\documentclass[reprint,amsmath,amssymb,aps,prl,noeprint,nolongbibliography,lengthcheck,superscriptaddress]{revtex4-2}
\usepackage[utf8]{inputenc}
\usepackage{graphicx}
\usepackage{amsmath}
\usepackage{xspace}
\usepackage{chemformula}

\newcommand{\supp}{Supplemental Material~\cite{SM}}

\begin{document}

\title{Machine-learning enabled optimization of atomic structures using atoms with fractional existence} 

\author{Casper Larsen}
\affiliation{CAMD, Department of Physics, Technical University of Denmark, Kongens Lyngby, Denmark}
\author{Sami Kaappa}
\affiliation{CAMD, Department of Physics, Technical University of Denmark, Kongens Lyngby, Denmark}
\affiliation{Computational Physics Laboratory, Tampere University, P.O. Box 692, FI-33014 Tampere, Finland}
\author{Andreas Lynge Vishart}
\affiliation{CatTheory, Department of Physics, Technical University of Denmark, Kongens Lyngby, Denmark}
\affiliation{ASM, Department of Energy Conversion and Storage, Technical University of Denmark, Kongens Lyngby, Denmark}
\author{Thomas Bligaard}
\affiliation{CatTheory, Department of Physics, Technical University of Denmark, Kongens Lyngby, Denmark}
\affiliation{ASM, Department of Energy Conversion and Storage, Technical University of Denmark, Kongens Lyngby, Denmark}
\author{Karsten Wedel Jacobsen}
\affiliation{CAMD, Department of Physics, Technical University of Denmark, Kongens Lyngby, Denmark}

\email{kwj@fysik.dtu.dk}
\date{\today}

\newlength{\figwidth}
\setlength{\figwidth}{0.95\columnwidth}
\newlength{\widefig}
\setlength{\widefig}{0.9\textwidth}

\begin{abstract}
We introduce a method for global optimization of the structure of atomic systems that uses additional atoms with fractional existence. The method allows for movement of atoms over long distances bypassing energy barriers encountered in the conventional position space. The method is based on Gaussian processes, where the extrapolation to fractional existence is performed with a vectorial fingerprint. The method is applied to clusters and two-dimensional systems, where the fractional existence variables are optimized while keeping the atomic positions fixed on a lattice. Simultaneous optimization of atomic coordinates and existence variables is demonstrated on copper clusters of varying size. The existence variables are shown to speed up the global optimization of large and particularly difficult-to-optimize clusters.
\end{abstract}

\maketitle

The atomic-scale structure is of critical relevance to the physical and chemical properties of materials and nanoparticles. In the low temperature limit, the most stable atomic configuration is found by minimizing the total energy, but the optimization problem is difficult because of many meta-stable states, and, in many cases, the total energy evaluations are computationally time consuming. 

To address these problems several algorithms of auto\-matized structure prediction have been proposed \cite{Zhang2020} including random searches \cite{randomsearch}, genetic searches \cite{EA2, sioclusters, Lysgaard2014, Jaeger2019}, basin hopping \cite{basinhopping} and particle swarm optimizations \cite{SGO}. Central to most of these methods is that they rely on carrying out large numbers of time-consuming calculations with density functional theory (DFT) or other quantum chemistry me\-thods. To circumvent the time-issue of DFT without compromising the accuracy of the calculations, Gaussian processes have shown effective in constructing surrogate potential e\-ner\-gy surfaces (PES) \cite{bartokGaussianApproximationPotentials2010, todorovicBayesianInferenceAtomistic2019}. These surfaces can be explored by random searching and updated by Bayesian search methods as demonstrated with the so-called GOFEE ('Global Optimization with First-principles Energy Expression') algorithm in Ref.~\onlinecite{Malthe2020efficient}. This methodology is generalized to include training on forces in the BEACON ('Bayesian Exploration of Atomic Configurations for OptimizatioN') code \cite{kaappa2021global}. In Ref.~\onlinecite{verner2022atomistic}, GOFEE is shown to decrease the number of energy evaluations necessary to find the global minimum by up to several orders of magnitude compared to traditional algorithms. Central to GOFEE/BEACON is the representation of atomic configurations by means of a fingerprint, which is invariant under translation, rotation, and inversion, and also under the permutation of atoms of the same chemical element.

It has been shown that the efficiency of random searching can be improved by inclusion of hyperdimensions \cite{pickard2019hyperspatial}. The extra dimensions make it possible to circumvent barriers in the usual configuration space. However, the energy function has to be defined for the extra hyper-dimensions. This can be done for some analytic interatomic potentials, but it is not clear how to do this in the case of potential energy surfaces based on quantum mechanical calculations.

An alternative way to increase the dimensionality of configuration space and circumvent barriers is to interpolate between chemical elements ('ICE') as implemented in the ICE-BEACON code \cite{kaappa2021atomic}. Here, additional dimensions are introduced so that an atom can be a fractional mixture of two chemical elements. The extension of the energy function to the extra dimensions is performed through a Gaussian process with a fingerprint, which allows for fractional chemical identities.

In this paper, we apply the idea of expanded dimensionality in a new way by introducing extra variables, which allow the atoms to have partial existence. The idea is that additional atoms of fractional existence can act as candidate sites for real atoms, allowing existence to be transferred from less to more favorable sites over arbitrarily long distances bypassing energy barriers in the conventional position space. Since some of the atoms end up with very little or no existence we shall refer to the additional atoms as ghost atoms, and we will refer to the approach as Ghost-BEACON.

In the model, a system with $N$ atoms is treated as a surrogate system with $N^\ast>N$ atoms, where every atom (with index $i$) is given a fractional existence $q_i\in[0,1]$ with the constraint that the fractions sum to the number of real atoms $\sum_i^{N^\ast} q_i = N$. The system is thus characterized by $3N^\ast$ spatial coordinates and $N^\ast$ existence variables. The existence variables are incorporated into a structural fingerprint with radial and angular parts that resemble the corresponding distribution functions. The radial part reads
\begin{equation}
\label{eq:fp_radial}
 { \rho}^R(r) = \sum_{\substack{i,j \\ i\neq j}} q_{i}q_{j}\frac{1}{r_{ij}^2}f_c(r_{ij}) \, e^{-|r-r_{ij}|^2/2\delta_R^2}
\end{equation}
where $r$ is the distance variable, $r_{ij}$ are the interatomic distances, $f_c$ a cutoff function, and $\delta_R$ a length parameter. The angular part has a similar form. (Please, see details of the machine learning model and the fingerprint in the \supp).

The radial fingerprint is in general quadratic in the existence variables. However, let us consider a situation where all atoms either fully exist ($q=1$) or are completely removed ($q=0$) except for two atoms, say numbers 1 and 2, whose distance is larger than the cutoff distance. In that case, the fingerprint becomes linear in $q_1$ and $q_2$. If we furthermore assume that the surroundings of the two atoms are identical, the transfer of existence from atom 2 to atom 1 ($q_2 = 1- q_1)$ leaves the fingerprint \emph{completely unchanged} during the transfer. This means that any machine-learning model based on the fingerprint shows no energy barrier for the process. This analysis also holds if the angular fingerprint is included. (Shown explicitly in \supp , Fig.~S1).

\begin{figure}[t!]
    \centering
    \includegraphics[scale=1.0]{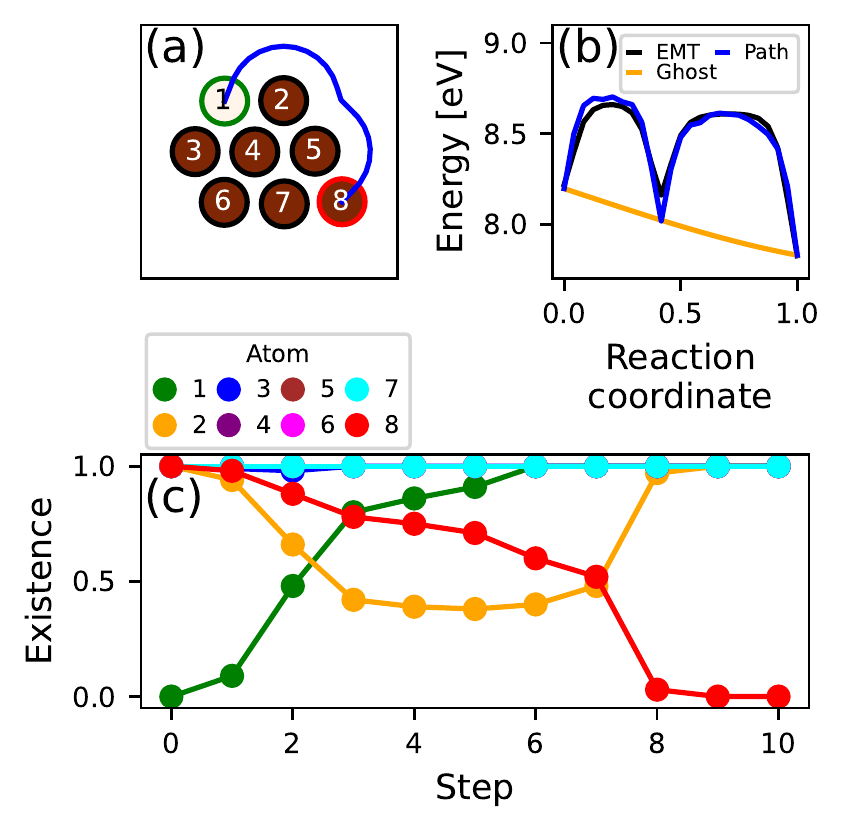}
    \caption{(a) The 2D test system with 8 atoms, labeled from 1 to 8. In this configuration, atom 1 is a ghost atom, and atoms 2-8 are real. The blue curve shows the real-space minimum-energy path, where atom 8 is moved to the empty site 1. (b) Different energy profiles while moving the atom from site 8 to site 1 in (a). The black curve shows the EMT energies along the minimum-energy path, and the blue curve shows surrogate energies along the same path. The yellow curve shows the energy profile in the case where no atoms are moved, but the existence is transferred from atom 8 to atom 1. 
    (c) The variation of the existence variables during the transfer of existence from atom 8 to 1.}
    \label{fig:illustration}
\end{figure}

To illustrate the removal of energy barriers further, we show in Fig.~\ref{fig:illustration}(a) a system with 7 copper atoms accompanied by a ghost atom with the energies calculated with an effective-medium-theory (EMT) interatomic potential \cite{jacobsenInteratomicInteractionsEffectivemedium1987,jacobsenSemiempiricalEffectiveMedium1996}.
We investigate the energy profile of moving an atom from a less favourable site (site 8) to a more favourable one (site 1) by following the trajectory shown in blue, which is the minimal-energy path found with a nudged-elastic-band (NEB) calculation \cite{millsQuantumThermalEffects1994, jonssonNudgedElasticBand1998}. We compare this motion to the alternative path of existence transfer allowed by the new existence variables. A Gaussian-process surrogate model is trained on 8 points along the NEB trajectory. The black curve in Fig.~\ref{fig:illustration}(b) shows the EMT energies along the NEB path, while the blue curve is the surrogate energy along the same path. The blue curve roughly matches the black one, as expected, showing two energy barriers in the energy landscape corresponding to atom 8 bypassing atoms 5 and 2. The yellow curve in Fig.~\ref{fig:illustration}(b) shows the energy during the transfer of existence from atom 8 to 1 with the reaction coordinate $q_1 = 1 - q_8$ and all other existence variables fixed. The energy is almost linear with no potential barrier which means that the transfer of the atom from site 8 to 1 is favoured and straightforward in the existence space.

Figure~\ref{fig:illustration}(c) visualizes the energy minimization process where initially $q_1 = 0$ and $q_i = 1$ for $i = 2, 3, \ldots, 8$. During the relaxation, the existence of atom 8 decreases while the existence of atom 1 increases. Interestingly, the process also involves atoms 2 and 3, which temporarily lose some of their existence. At the end of the relaxation, the existence has been completely transferred from atom 8 to atom 1. 

\begin{figure}[t!]
    \centering
    \includegraphics[scale=1]{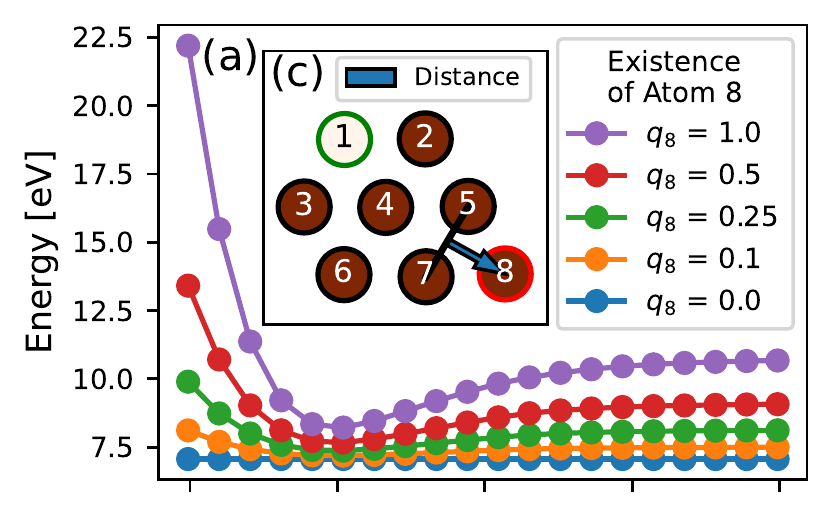}\vspace{-0.8em}
    \hspace*{-0.5em}\includegraphics[scale=1]{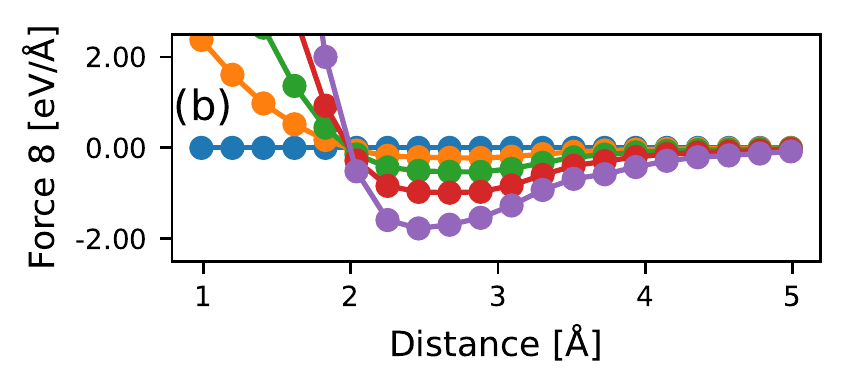}
    \caption{(a) Energy curve and (ba) force curve of copper atom 8 as a function of the distance between copper atom 8 and the remaining cluster along the direction of the blue arrow depicted in (c) for different existence fractions of atom 8. Training is done with EMT on 10 different distances of atom 8. All existence not carried in atom 8 is placed in atom 1 ($q_1 = 1- q_8$). The energy curves are seen to exhibit a minimum at approximately the same distance.}
    \label{fig:forces}
\end{figure}

We further illustrate the property of the PES when varying the existence variables in Fig.~\ref{fig:forces}. Atom 8 is now moved along the indicated linear path in Fig.~\ref{fig:forces}(c) when having different amounts of existence $q_8$, where the remaining existence is taken up by atom 1, $q_1 = 1-q_8$. Atom 8 is seen to be more weakly interacting with the rest of the cluster when its existence is reduced, but the bonding distance remains essentially the same. This means that an atom with a small existence will tend to position itself at similar geometries as real atoms making the transfer of existence more relevant. However, the figure also shows that an atom with vanishing existence does not interact. This also follows from the fact that such an atom does not contribute to the fingerprint. Atoms with zero existence can therefore float freely around making it unlikely that they take part in optimization. For efficient structure optimizations, it is therefore necessary to introduce a lower bound for the existence variables and consequently increase the total existence. 

It should be noted that the extension of the machine learning model to the fractional existence space is an extrapolation that cannot be controlled by the addition of data points. The quality of the model therefore depends strongly on the way the existence fractions are included in the fingerprint and the choice of hyperparameters for the machine learning model.

\begin{figure}[ht]
    \centering
    \includegraphics[scale=0.92]{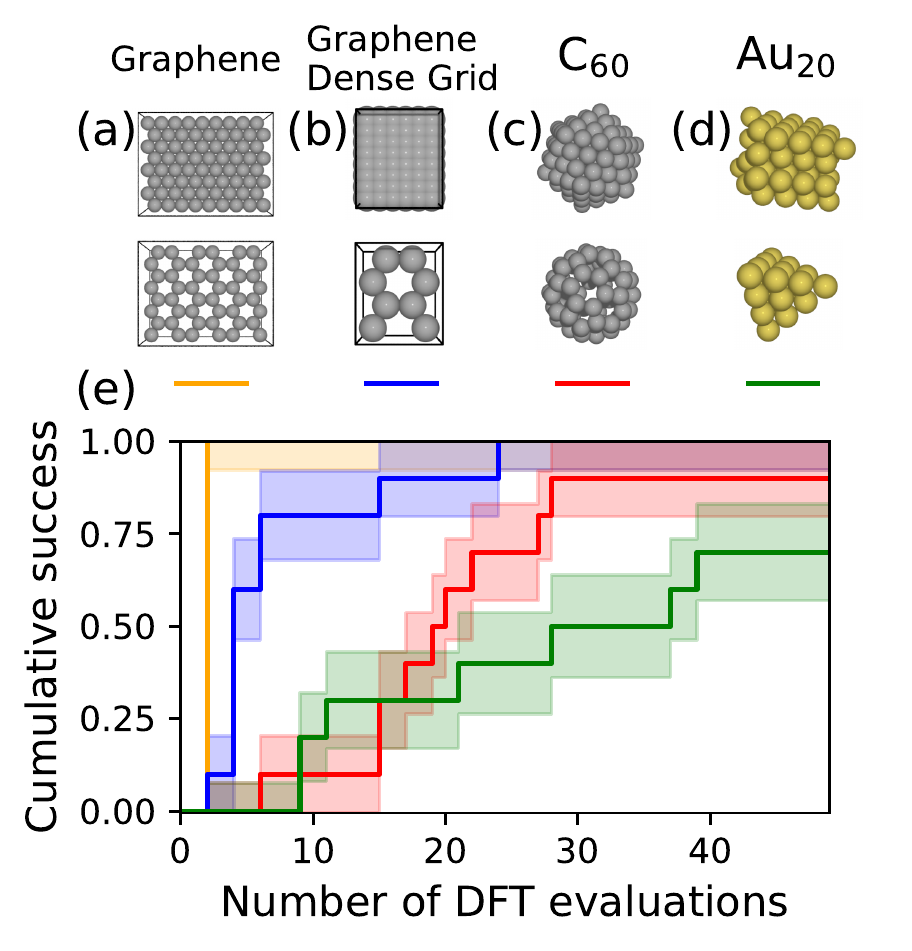}
    \caption{(a-d) Atomic grids (top) and global minimum energy structures (bottom) of (a) carbon (48 atoms) on a periodic triangular lattice (72 atoms), (b) carbon (8 atoms) on a dense rectangular lattice (48 atoms total), (c) C$_{60}$ on a 147 atoms icosahedral grid, and (d) Au$_{20}$ on a 64 atoms fcc grid. (e) Success curves for finding the global minimum energy structure for each setup shown in (a-d). Only the existence variables are optimized while keeping the atomic positions fixed on the grid. The uncertainties are Bayesian estimates.}
    \label{fig:grid}
\end{figure}

We now turn to structural optimizations where the energies and forces are based on DFT. The DFT calculations are performed using GPAW \cite{mortensenRealspaceGridImplementation2005, enkovaaraElectronicStructureCalculations2010} and the Atomic Simulation Environment \cite{ase-paper, ase}. We apply the Perdew-Burke-Ernzerhof \cite{PBE} exchange-correlation functional. The plane wave cutoff is 700 eV and the Fermi temperature is 0.1 eV. Only the $\Gamma$-point is used for k-point sampling except for graphene on a dense grid (Fig.~\ref{fig:grid}) where (3,2,1) k-points are used. When performing relaxations with DFT, we use as convergence criterion that all atomic forces are smaller than 0.01 eV/Å.

The optimization algorithm is similar to the one of ICE-BEACON but with existence variables instead of chemical element interpolation: given a database of structures with DFT calculated energies and forces, a surrogate PES is constructed using a Gaussian process where the structures are described by the fingerprint. All systems in the database have $N$ atoms, but the surrogate model can be used to make predictions for systems with $N^\ast$ atoms with fractional existence. The surrogate PES is explored with random searching, that is with 40 local relaxations based on random initial configurations. The relaxations can be performed in either the atomic coordinates or the fractional existence variables, or both. If the existence variables take on fractional values after relaxation, the $N$ largest fractions are set to 1, and the remaining to 0. The relaxed structures are evaluated with an acquisition function using the predicted energy and its uncertainty, and the structure with the lowest value is added to the DFT database. This procedure is iteratively repeated keeping track of the low energy structures obtained. The full simulation procedure is repeated to obtain statistics of the performance. Details of the algorithm including the computational parameters can be found in \supp.

We first consider some examples where the atomic positions are fixed and where only the existence variables are optimized. Fig~\ref{fig:grid}(a)-(d) show four different systems, which are 
(a) a single layer of carbon atoms on a periodic triangular lattice with an equilibrium interatomic distance of 1.42 Å corresponding to the one of graphene. The system contains a total of 72 atoms with 48 real atoms, which is the number of atoms corresponding to a layer of graphene. (b) A dense layer of carbon atoms on a periodic rectangular grid with interatomic distance $a = 0.710$ Å in one direction and $0.5\sqrt{3} a$ in the other direction. The total number of atoms is 48 with 8 real atoms again corresponding to the density of graphene. (c) An icosahedron of carbon atoms with 147 atoms in total and 60 real atoms with an interatomic distance of 1.44 Å between atoms belonging to the same icosahedral layer roughly agreeing with the bond lengths for a Bucky ball. (d) A cluster of fcc gold containing a total of 64 atoms and 20 real atoms.

Each optimization has an initial training set of two random sets of existence variables: one where the atoms are chosen by random and one where the atoms are chosen by random but so that the final structure is connected.
The obtained minimum-energy structures for the four systems are shown in the lower panel of Fig.~\ref{fig:grid}(a)-(d)  The minimum-energy structure for (a) and (b) is a graphene layer, for (c) it is a \ch{C60} bucky ball, and for (d) it is the tetrahedral \ch{Au20} cluster \cite{liAu20Tetrahedral2003}.
The statistics of the optimizations are shown in the success curves in Fig~\ref{fig:grid}(e). In all four cases 10 independent simulations have been performed, and the success curves show the fraction of simulations, which have found the lowest-energy structure as a function of the number of DFT calculations being performed.

The algorithm succeeds in finding the global optimum within 50 DFT calculations in 10/10 runs for both grid types of graphene and in 9/10 and 7/10 attempts for C$_{60}$ and Au$_{20}$, respectively. Finding the structure of graphene on the standard triangular lattice proved to be a particularly easy task for the algorithm, which is probably due to the high degree of regularity of the grid and due to the high $N/N^\ast$ ratio as compared to the problem of Au$_{20}$, for example. 

\begin{figure}[ht]
    \centering
    \includegraphics[scale=0.91]{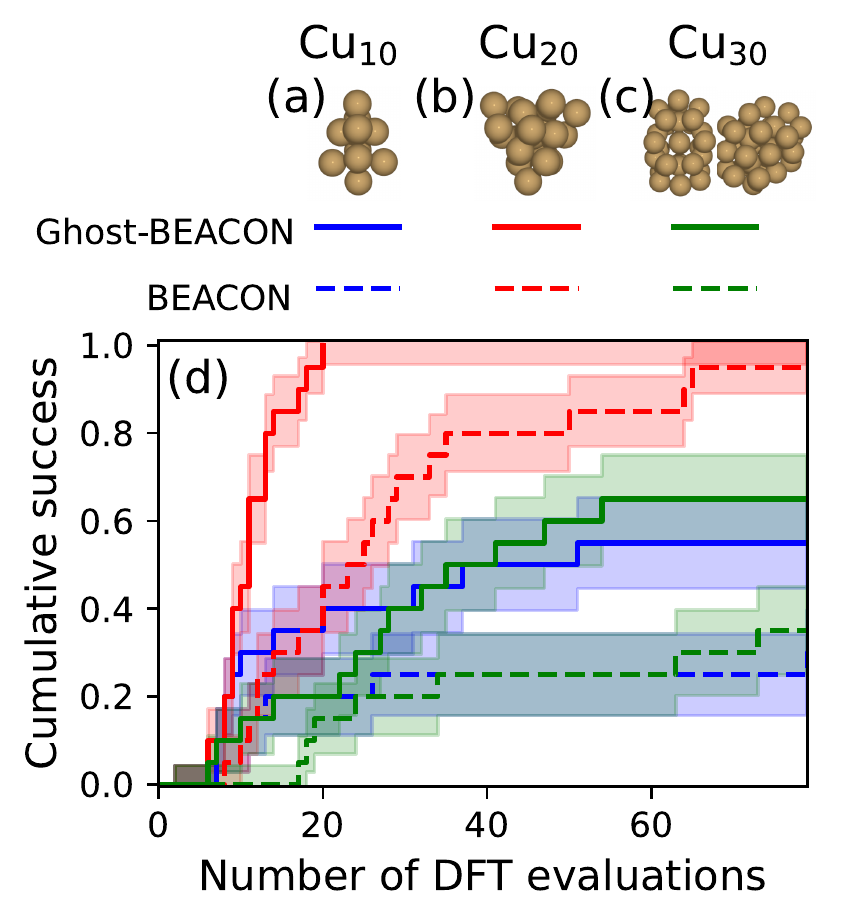}
    \caption{(a-b) Global minimum structures of Cu$_{10}$ and Cu$_{20}$ and(c) the two lowest energy minima of Cu$_{30}$ being so close in energy that they are almost inseparable. (d) Success curves of 20 independent runs of each 80 DFT-calculations without ghost atoms (BEACON) and with ghost atoms constituting 1/3 of the total number of atoms (Ghost-BEACON) for optimization of Cu$_{10}$ (5 extra atoms), Cu$_{20}$ (10 extra atoms) and Cu$_{30}$ (15 extra atoms). Each iteration of the BEACON cycle was based on 40 surrogate relaxations. Each run had an initial training set of 2 random structures.}
    \label{fig:cluster}
\end{figure}

The method also allows for simultaneous optimization of atomic coordinates and existence fractions as we shall now illustrate with copper clusters of varying size. We compare the performance of BEACON, which optimizes in only the configuration space of atomic coordinates, and the present approach, Ghost-BEACON, which optimizes in both configuration space and existence variables. We consider clusters of sizes 10, 20, and 30 atoms and in each case we add 50\% ghost atoms and perform 20 independent simulations. The resulting minimum-energy structures are shown in Fig.~\ref{fig:cluster} together with the success curves, where success is declared when a structure is within 0.1 eV of the lowest energy encountered across all runs of a give cluster size. Further analysis shows that the declared successful structures for \ch{Cu10} are all identical, while in the case of \ch{Cu20} two distinct structures are identified. In the case of \ch{Cu30} several structures have low energies, most of them slight alterations of the structures shown in (c).

We first note that the number of DFT calculations necessary to determine low energy structures does not vary monotonically with cluster size. The \ch{Cu10} cluster requires considerably more computational effort than \ch{Cu20}. This might seem surprising as the number of variables to consider in the optimization of course increases with cluster size. However, it should be recalled that we are doing random searching on the surrogate PES (with or without the existence variables) starting from random initial configurations, and the basin of attraction for the different local minima might vary substantially. This is the case for \ch{Cu10}, where the 3rd lowest energy structure is found more frequently than the ground state. (Shown with success curves in \supp Fig.~S2).

The presence of ghost atoms is seen to improve the searches considerably, in particular in the cases where BEACON does not easily identify the ground state.

The structures of Fig.~\ref{fig:cluster}(a-c) are different from the ones found using empirical potentials or tight binding molecular dynamics \cite{boyukata2008structural, doye1998global, kabir2004copper}. They are also different and lower in energy than the structures found using DFT in Ref.~\onlinecite{rangel2021nonconventional} as verified by relaxing all candidate structures with DFT. 

\begin{figure}[ht]
    \centering
    \includegraphics[width=1.0\columnwidth]{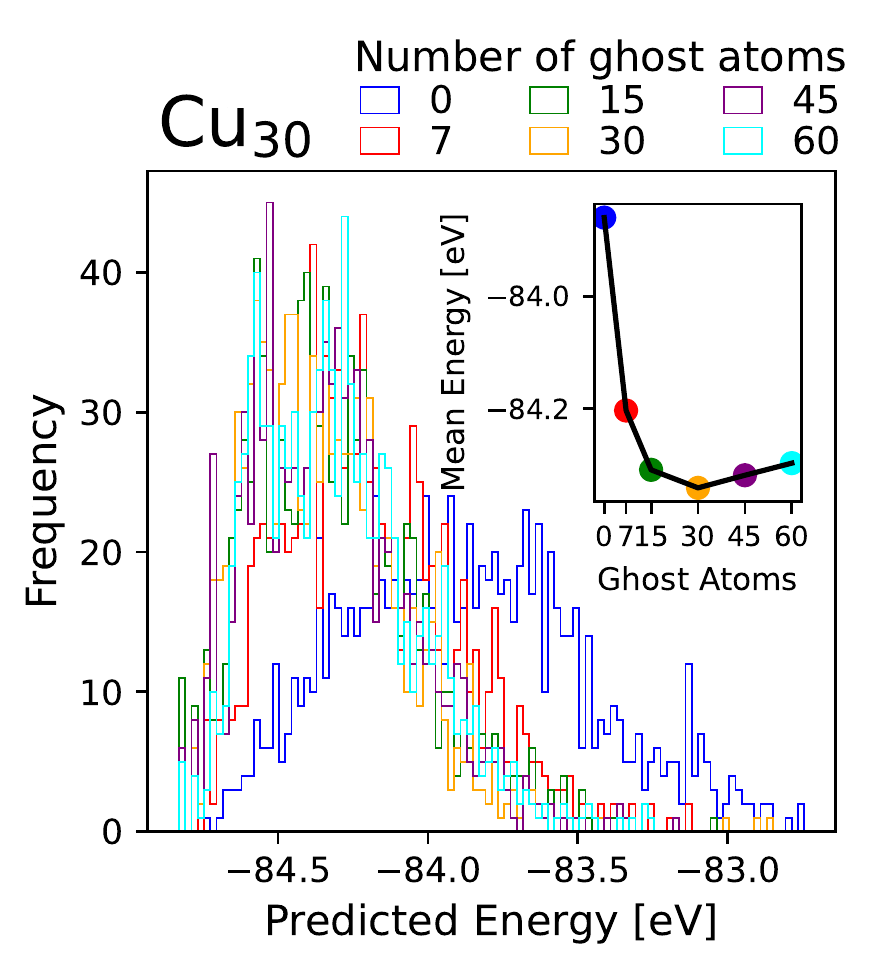}
    \caption{Distribution of the energies obtained by 1000 relaxations on a surrogate PES for \ch{Cu30}. The inset shows the variation of the average energies as a function of the number of ghost atoms.}
    \label{fig:histograms}
\end{figure}

The main function of the ghost atoms is to open new relaxation pathways as discussed above. To analyze this more, we construct a surrogate PES for \ch{Cu30} from a training set consisting of 151 configurations including some of the identified low-energy structures. We perform 1000 relaxations on the potential energy surface from random initial configurations for different choices of ghost atoms. The distributions of the obtained relaxed surrogate energies are shown in Fig.~\ref{fig:histograms}. Without any ghost atoms (the blue curve) we get the result that is obtained with BEACON. We see that when ghost atoms are introduced, the distribution is shifted to lower energies as an indication that the relaxations are not trapped as much in higher-lying local minima as is the case for BEACON. The inset in the figure shows the average energies of the distributions. Clearly the main effect comes from introducing just a few ghost atoms into the system, and the effect quickly levels off with the number of ghost atoms. The fact that rather few ghost atoms improve the efficiency is also seen for Cu$_{10}$ and Cu$_{20}$ and is also observed in the success curves (\supp Figs.~S3 and S4).

Several modifications and extensions of the approach presented here are possible. It should be straightforward to combine the method with the ICE-approach. Each atom $i$ would then carry a set of variables $q_i^A\in [0,1]$, where A indicates the chemical element. The total existence of the atom would then be given by $q_i = \sum_A q_i^A\in[0,1]$ with the constraint that the number of atoms $N_A$ of element A is $N_A = \sum_i^{N^\ast} q_i^A$.

The example with graphene on a dense grid points to the possibility of restricting the atomic positions to a finely spaced grid and then only optimize the existence variables. However, this will require the treatment of very many atoms (one per grid point), which is not feasible with the current fingerprint.

In the present implementation, the sum of the existence variables is constrained to be the number of real atoms in the system. However, one could easily generalize this to treat open systems with a variable number of atoms controlled by a chemical potential. This would just correspond to a Lagrange-multiplier implementation of the constraint.

We acknowledge support from the VILLUM Center for Science of Sustainable Fuels and Chemicals, which is funded by the VILLUM Fonden research grant (9455).

\bibliography{bibliography}

\end{document}


\title{Supplemental Material for:\\Machine-learning enabled optimization of atomic structures using atoms with fractional existence} 

\author{Casper Larsen}
\affiliation{CAMD, Department of Physics, Technical University of Denmark, Kongens Lyngby, Denmark}
\author{Sami Kaappa}
\affiliation{CAMD, Department of Physics, Technical University of Denmark, Kongens Lyngby, Denmark}
\affiliation{Computational Physics Laboratory, Tampere University, P.O. Box 692, FI-33014 Tampere, Finland}
\author{Andreas Lynge Vishart}
\affiliation{CatTheory, Department of Physics, Technical University of Denmark, Kongens Lyngby, Denmark}
\affiliation{ASM, Department of Energy Conversion and Storage, Technical University of Denmark, Kongens Lyngby, Denmark}
\author{Thomas Bligaard}
\affiliation{CatTheory, Department of Physics, Technical University of Denmark, Kongens Lyngby, Denmark}
\affiliation{ASM, Department of Energy Conversion and Storage, Technical University of Denmark, Kongens Lyngby, Denmark}
\author{Karsten Wedel Jacobsen}
\affiliation{CAMD, Department of Physics, Technical University of Denmark, Kongens Lyngby, Denmark}

\date{\today}

\maketitle

\section{Machine learning model}
\subsection{Fingerprint}

\noindent The fingerprint is based on the one used in BEACON \cite{kaappa2021global} with the inclusion of existence fractions $q_i\in[0,1]$ for each atom $i$. It is similar to the one used in ICE-BEACON \cite{kaappa2021atomic} with the difference that in ICE-BEACON all atoms have both a fraction $q_{i,A}$ of element A, and a fraction for element B that satisfies $q_{i,B}=1-q_{i,A}$, while here $q_i$ denotes the total existence of the atom.

The fingerprint is denoted by $\rho(\mathbf{x},Q)$, where $\mathbf{x}$ is the full set of Cartesian coordinates and $Q$ is the full set of existence fractions. $\rho(\mathbf{x},Q)$ is divided into a radial part, $\rho^R(r; {\mathbf x}, Q)$ and an angular part, $\rho^\alpha(\theta; \mathbf x, Q)$, which for a single-element system is given by:

\begin{equation}
\label{eq:fp_radial}
 { \rho}^R(r; {\mathbf x}, Q) = \sum_{\substack{i,j \\ i\neq j}} q_{i}q_{j}\frac{1}{r_{ij}^2}f_c(r_{ij}; R_c^R) \, e^{-|r-r_{ij}|^2/2\delta_R^2}
\end{equation}
\begin{equation}
\label{eq:fp_angular}
    { \rho}^\alpha(\theta; \mathbf x, Q) = \sum_{\substack{ i,j,k \\ i\neq j \neq k }}  \Big( q_{i} q_{j} q_{k} f_c(r_{ij}; R_c^\alpha)f_c(r_{jk}; R_c^\alpha) \cdot 
    e^{-|\theta-\theta_{ijk}|^2/2\delta_\alpha^2}\Big)
\end{equation}
\begin{equation}
\label{eq:cutoff}
    f_c(r_{ij};R_c)=\begin{cases}
    1-(1+\gamma)\big( \frac{r_{ij}}{R_c} \big)^\gamma+\gamma\big(\frac{r_{ij}}{R_c}\big)^{1+\gamma}& \text{if } r_{ij}\leq R_c \\
    0& \text{if }r_{ij}>R_c
\end{cases}
\end{equation}

where the indices $i$, $j$, and $k$ run over all atoms. Here $r_{ij}$ is the distance between atoms $i$ and $j$, $\theta_{ijk}$ is the angle between atoms $i$, $j$ and $k$, and $f_c$ is a smooth cutoff function going to zero at the radial and angular cutoff radii $R_c^R$ and $R_c^\alpha$, respectively. $\gamma$ is a parameter set to 2. Hence $\rho^R$ describes a sum over all pairs of atoms whereas $\rho^\alpha$ describes a sum over all triplets. The full fingerprint $\rho( \mathbf{x},Q)$ is created by concatenating $\rho^R$ and $\rho^\alpha$. 

In Eqs.~\ref{eq:fp_radial} and \ref{eq:fp_angular}, the values for $R_c^R$ and $R_c^\alpha$ are fixed for a given system but scaled with the covalent radius $r_{cov}$ of the element as $R_c^R=5 r_{cov}$ and $R_c^\alpha=3 r_{cov}$. The constants $\delta_R=0.4$ Å and $\delta_\alpha=0.4$ rad are identical for all systems.  

\subsection{Gaussian process in the Ghost-BEACON framework}

\noindent Following the notation for the fingerprint, the energies and forces, $\mu=(E, -F)$, are calculated with the standard expression for a Gaussian Process \cite{williams2006gaussian,poloczek2017gradients}:

\begin{equation}
\label{eq:predict}
    \mu(\mathbf x, Q) = \mu_{p}(\mathbf x, Q) + K(\rho[\mathbf x, Q], P)C(P, P)^{-1}(y - \mu_p(X))
\end{equation}
where $\mu_p(\mathbf x, Q)$ and $\rho( \mathbf{x},Q)$ are the prior mean and the fingerprint, respectively, $K$ and $C$ are the covariance matrix without and with regularization, $P$ a matrix containing the training data fingerprints, $y$ the training data targets and $\mu_p(X)$ the prior function applied to all structures in the training data. The uncertainty of the predicted energy is given by: 

\begin{equation}
\label{eq:uncertainty}
    \Sigma(\mathbf x, Q) = \Big\{\Tilde{K}(\rho[\mathbf{x},Q],\rho[\mathbf{x},Q])-K(\rho[\mathbf{x},Q],P)C(P,P)^{-1}K(P,\rho[\mathbf{x},Q])\Big\}^{1/2} ,
\end{equation}
where $\Tilde{K}(\rho(\mathbf{x},Q),\rho(\mathbf{x},Q))$ represents the covariance matrix for the fingerprint.

The applied kernel function for the covariance matrices is a squared exponential kernel function (SE). The SE uses a prefactor ($\sigma^2$) and one length-scale (l) hyperparameters (the routine for optimization of the hyperparameters is described below). The covariance matrix between two atomic configurations has three components\cite{kaappa2021global,koistinen2017Nudged}. The first components are the covariances between energies ($k$), the second are the covariances between energies and forces ($\nabla_{i}k$), and the third component are the covariances between forces ($\nabla_{i}\nabla_{j}k$). $\nabla_{i}$ is the gradient operator with respect to the Cartesian coordinates $\mathbf{x}_i$. The covariance matrix is written as
\begin{equation}
    K(\rho_1,\rho_2) = 
     \begin{bmatrix}
           k(\rho_1,\rho_2) &
           (\nabla_2 k(\rho_1,\rho_2))^{\top} \\
           \nabla_1 k(\rho_1,\rho_2) &
           \nabla_1 (\nabla_2 k(\rho_1,\rho_2))^{\top} 
   \end{bmatrix}.
\end{equation}
We observe from Eq.~\ref{eq:predict} that $K$ and $\mu_{p}(\mathbf{x},Q)$ are the only terms including the existence fractions. Details about the construction of $K$, $C$, and $y$ are reported in Ref.~\cite{kaappa2021global}. Keeping the order of all terms but simplifying the notation, we can rewrite Eq.~\ref{eq:predict} as 
\begin{equation}
\label{eq:predict_simple}
    \mu = \mu_{p,\mathbf{x}} + K C^{-1}(y - \mu_{p,X}).
\end{equation}
If we denote the number of atoms by $N$, the number of elements per data point will be $F=1+3N$ for one energy and $3N$ force components. If we further denote the number of structures in our training data $D$, the full training data will include $DF$ features. Keeping the order of terms in Eq.~\ref{eq:predict_simple}, we have the following dimensions:
\begin{equation}
\label{eq:predict_simple_dim}
    [F] = [F] + [F \times DF] [DF \times DF][DF]
\end{equation}
which is the standard scenario for a Gaussian process. When predicting features on a structure with $N^\ast$ atoms (comprising the real and the ghost atoms), the amount of predicted features becomes $G=1+3N^\ast$, but the number of features on all structures in the training data is still $F$ and hence Eq.~\ref{eq:predict_simple_dim} becomes
\begin{equation}
\label{eq:predict_simple_dim_ghost}
    [G] = [G] + [G \times DF] [DF \times DF][DF]
\end{equation}

\subsection{Prior function}
For the simultaneous optimization of positions and existence fractions of Figs.~3 and 4 in the main text, a repulsive prior modified to include the existence fractions is used \cite{Malthe2020efficient,kaappa2021global}: 
\begin{equation}
\label{eq:repulsive}
    \mu_{p}(\mathbf x, Q)=\mu_c+\sum_{\substack{i,j \\ i\neq j \\ r_{ij}<2R} }q_iq_j\Big(\frac{2\sigma_p \Tilde{r}_{cov}}{r_{ij}}\Big)^{12},
\end{equation}
where $\sigma_p$ is a repulsive constant set to 0.4 and $\Tilde{r}_{cov}$ is an atomic radius set to be $0.8r_{cov}$ of the element and $\mu_c$ is a constant prior. This prior is chosen to disfavor atoms with overlapping atomic radii, but in such a way that low existence atoms do not interfere with the clustering of high existence atoms. For all other simulations the prior is set to a constant value $\mu_p=\mu_c$ which is updated throughout the run.

\subsection{Acquisition function}
\noindent We use the acquisition function $f$ for a structure $\mathbf x$ given by $f(\mathbf x)=\mu(\mathbf x) - \kappa \Sigma(\mathbf x)$,
where $\kappa=2$ is a constant while $\mu(\mathbf x)$ and $\Sigma(\mathbf x)$ are the predicted energy and uncertainty of Eq.~\ref{eq:predict} and Eq.~\ref{eq:uncertainty} \cite{Malthe2020efficient,kaappa2021global}.
The dependency on $Q$ is omitted as the acquisition function is always evaluated on structures without ghost atoms.

The acquisition function is used to select which of the relaxed structures to include in the DFT database. However, sometimes the relaxations mostly reproduce an already investigated structure. It is therefore an advantage to discard structures that are closer than a certain distance, $d_\textrm{fp}$, in fingerprint space from already known structures. 
For the optimization of both atomic coordinates and existence values we set $d_\textrm{fp}=5$. For the optimization on a grid $d_\textrm{fp}$ was set to a small value to exclude already visited structures without disqualifying any other structures.

\subsection{Robust determination of hyperparameters}\label{subsec:hyper_opt}
\noindent During the BEACON and Ghost-BEACON runs, the hyperparameters are updated by using the maximum a posteriori probability (MAP) for the hyperparameters given the training data, $p(l,\sigma_r,\sigma|y)$. 
A uniform prior is considered for the prefactor, $\sigma$, whereas a log-normal prior distribution is used for the length-scale hyperparameter, $l$, as explained in the section below.  The noise, $\sigma_n$, is set by a relative noise, $\sigma_r^2=\frac{\sigma_n^2}{\sigma^2}$.

The MAP is calculated by using the analytical solution of the prefactor, $\sigma_{MLE}^2$, from maximizing the posterior distribution:
\begin{equation}
    \sigma^2_{\text{MLE}}=\frac{1}{DF}(y-\mu_{p})^{\top} C_0^{-1} (y-\mu_{p}) \label{eq:prefactor_mle}
\end{equation}
where $C_0(P,P)=K_0(P,P)+\sigma_r^2I$ is the covariance matrix of the training data without the prefactor and a relative-noise. 
$K_0(P,P)$ denotes the covariance matrix of the training data without the prefactor and noise. The same relative-noise is used for energy and force contributions. The log-posterior distribution, $\mathit{LP}$, is: 
\begin{equation}
    \mathit{LP}(l,\sigma_r,y) \propto \mathit{MLL}(l,\sigma_r,y) + \ln{\left(p(l) \right)} \label{eq:lp}
\end{equation}
where the $\mathit{MLL}$ is the maximum log-likelihood with respect to the prefactor hyperparameter. The $\mathit{MLL}$ is expressed as:
\begin{align}
    \mathit{MLL}
    =& \frac{-1}{2 } \left(DF + \ln{\left(|C_0| \right)} + DF \ln{\left(\frac{1}{DF}(y-\mu_{p})^{\top} C_0^{-1} (y-\mu_{p}) \right)} + DF\ln{\left(2\pi \right)} \right) \notag \\
    =& \frac{-1}{2 } \left(DF + \sum_{i=1}^{DF}{ \ln{\left( [\Lambda]_{ii}+\sigma_r^2 \right)}} + DF \ln{\left(\frac{1}{DF}\sum_{i=1}^{DF} {\frac{[E^{\top}(y-\mu_{p})]_i^2}{[\Lambda]_{ii}+\sigma_r^2}} \right)} + DF\ln{\left(2\pi \right)} \right)
    \label{eq:ll_mle}
\end{align}
where $E$ is the eigenvectors and $\Lambda$ is the diagonal matrix with the eigenvalues of the covariance matrix without prefactor and relative-noise hyperparameters, $K_0(P,P)=E\Lambda E^{\top}$. All relative-noise hyperparameter values can be searched from a single eigendecomposition. A small noise is added to the covariance matrix to ensure it is invertible. However, a fixed relative-noise of 0.001 is used to avoid the maximum likelihood values that corresponds to the overfitting models.

A uniform grid with a spacing of 0.1 in the log-space of the length-scale hyperparameter is constructed from the mean nearest neighbour to 100 times the maximum Euclidean distance in the fingerprint space. All intervals surrounding a maxima of the log-posterior can be identified by using finite difference on the grid. Afterwards, a golden-section search is performed for all intervals containing a maxima.

The grid search method finds the global maxima of the log-posterior distribution under the constraints if the grid spacing is finer than the length of the basin of attraction.

The prior mean constant is optimized from the maximum likelihood \cite{kaappa2021global} under the constraint that it must be greater than or equal to the average between the smallest and the mean energies of the training data as:
\begin{equation}
    \mu_{p}=\begin{cases}
    \frac{(\textrm{min}(Energy)+\textrm{mean}(Energy))}{2} & \text{if } \mu_{p}<\textrm{min}(Energy) \\
    \frac{\mathbf{U}^{\top}C(P,P)^{-1}y)}{\mathbf{U}^{\top}C(P,P)^{-1}\mathbf{U}} & \text{otherwise} \\
\end{cases}
\end{equation}
where $\mathbf{U}$ is a vector with the length of $DF$ and has $U_i=1$ for energy components and $U_i=0$ for force components. 

The prior distribution and constrained interval of the length-scale hyperparameter improves the model quality at small data sets at the beginning of a run, where the model could be likely to either overfit (short length-scales with low noise) or underfit (very large length-scales with high noise). At the beginning of a run, the length scale is set to 2.5 times the maximal distance in fingerprint space. The prefactor, noise and prior are updated at every BEACON cycle, whereas the length scale is updated every fifth cycle. 

\subsection{Prior distribution of the length scale}

A prior distribution of the length scale is introduced to hinder the algorithm in over-fitting for small data sets and because it is observed that a longer length scale improves the interpolation in existence space. The length scale prior is defined as a log-normal distribution, i.e. a normal distribution in the logarithmic space:

\begin{equation}
P(l)=\frac{1}{l\sigma_{LN}\sqrt{2\pi}}\exp\Big(-\frac{(\ln(l)-\mu_{LN})^2}{2\sigma_{LN}^2}   \Big),
    \label{eq:lognormal}
\end{equation}
where $\mu_{LN}$ and $\sigma_{LN}^2$ are the mean and variance in the logarithmic space.  

A simple estimate of the length is $l_0 = 0.5(\textrm{mean}(\Delta_{FP})+\textrm{max}( \Delta_{FP}))$, where $\Delta_{FP}$ are all the Euclidian  distances between any two fingerprints. We set the parameters $\mu_{LN}$ and $\sigma_{LN}$ using $\textrm{mode}(l) = \exp(\mu_{LN} - \sigma_{LN}^2)= l_0$ and take $\sigma_{LN} = 0.75$.

For the Gaussian processes for EMT-evaluated Cu structures, which are fitted to only few data points, we simply keep the initial estimate of the length as 2.5 times the maximal distance in the fingerprint space. 

\section{Algorithmic details and computational parameters}

\subsection{Random structure generator}

\noindent In this study, all random configurations not placed on a grid are set up using a cubic box with a volume which is five times the sum of the volumes of atomic spheres with radii equal to the covalent atomic radii of the elements. The atoms initially placed randomly in the box are then repelled until all atom centers are at least $1.6 r_{cov}$  away from each other. 7.5 Å of vacuum is then added around the structure to complete the unit cell. This procedure ensures a similar initial atomic packing fraction independent on the number of atoms in the BEACON/Ghost-BEACON runs.

\subsection{Random fraction generator}
The random sampling of the initial existence values is done using the Dirichlet-Rescale algorithm \cite{GriffinRTSS2020, david_griffin_2020_4118059}. This allows for a uniform distribution of the existence values satisfying the constraints $q_i\in [q_\textrm{min},1]$ and $\sum_i q_i = N+(N^\ast-N)q_\textrm{min}$, where $0\leq q_\textrm{min} < 1$ is the lower existence bound.

\subsection{Surrogate surface relaxations}
The relaxations on the surrogate potential energy surface are performed using sequential least squares programming \cite{slsqp} as implemented in the SCIPY package \cite{scipy}. This allows for efficient gradient-driven optimization under the inequality constraint that all atoms have an existence value between $q_\mathrm{min}$ and $1$ as well as the equality constraint for the total amount of existence.

While optimizing the coordinates and the existence fractions simultaneously, the existence of an atom might fall to zero, effectively removing its interactions with the rest of the system. To counteract this unwanted effect in the algorithm and to proceed with the most efficient optimization, a lower limit to the existence is introduced, and the following procedure is adopted:\\
1) Initialize a system of random atomic positions and existence fractions between $q_\textrm{min}^\textrm{init}$ $(>0)$ and 1 with a total existence of $N + (N^\ast -N) q_\textrm{min}^\textrm{init}$. \\
2) Relax the system on the surrogate PES for $n_{relax}$ steps. \\
3) Decrease the lower limit in $n_D$ steps of $q_\textrm{min}^\textrm{init}/n_D$ and, at each level, perform a relaxation with $n_d$ steps.\\
4) Relax the system for $n_{p}$ steps with all existence variables fixed to 0 or 1 to effectively remove the ghost atoms.\\\\
The relaxations are terminated if all predicted forces are below 0.001 eV/Å. The low value is picked to counteract underestimation of forces in regions of large uncertainty on the potential energy surrogate surface. In this paper, the simultaneous relaxations of existence and positions are done with $n_{relax}=200$, $q_\textrm{min}^\textrm{init}=0.05$, $n_D=5$, $n_d=20$, and $n_p=100$. 

The calculations on a grid, where only the existence variables are optimized, do not require a lower boundary. All non-ghost BEACON runs are performed with $n_{relax}=400$ with $N^\ast=N$ and all fractions fixed to 1.

\subsection{Declaration of success}

\noindent Except for Fig.~4 in the main paper, a success is registered once a structure satisfies the correct nearest neighbor distribution for all atoms in the cluster as compared to the global minimum. This procedure is chosen to identify structures belonging to the correct basin. 

\subsection{Calculation of success curve uncertainty}

To calculate the uncertainty of the success curves of the paper, a Bayesian approach was followed. A success curve composed of $W$ independent runs can for a given number of DFT calculations be seen as a binary outcome of $n$ successes and $m$ failures such that the total number of attempts is always $W=n+m$.
Using Bayes theorem with a uniform prior, the posterior probability of the chance of success $p_{success}$ becomes a Beta distribution $B(p|\alpha=n+1,\beta=m+1)$. We use the mode of this distribution $\textrm{mode}(p_{success})=n/(n+m)$ as the value of the success curve. For the uncertainty,  we use the square root of the variance
\begin{equation}
\sqrt{\textrm{var}(p_{success})}=\sqrt{\frac{(n+1)(m+1)}{(n+m+2)^2(n+m+3)}}.
\end{equation}

\clearpage

\section{Supplementary data}

\begin{figure*}[ht]
    \centering
    \includegraphics[scale=0.90]{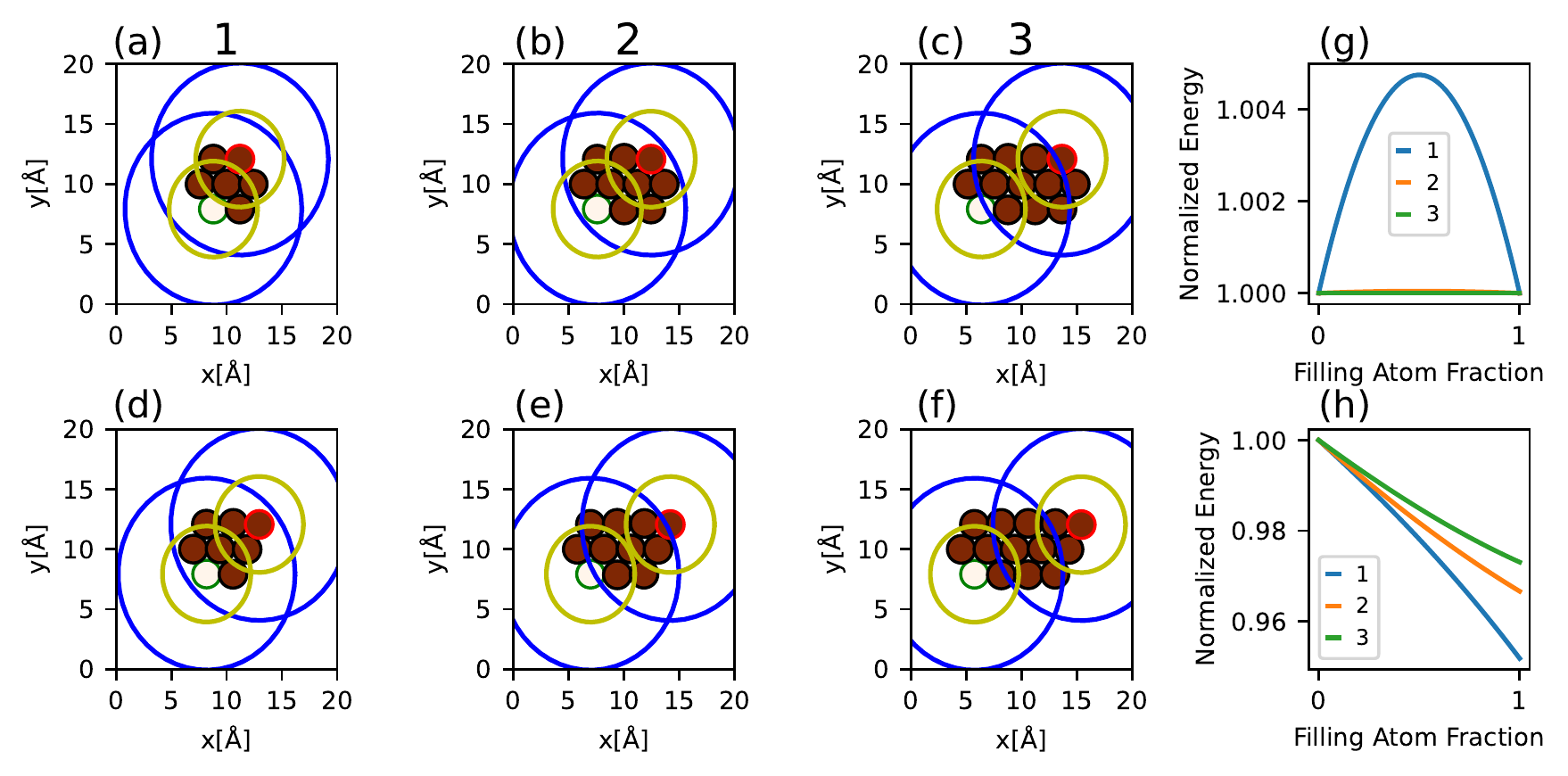}
    \caption{Illustration of the point that existence transfer between atoms far away from each other does not involve energy barriers. The clusters are similar to the one in Fig.~1 of the main paper. The clusters have different distances between the two atoms indicated with red and green edges. In the upper row (a-c), the two atoms occupy identical sites while in the lower row (d-f) they are different. Blue and yellow circles indicate the radial and angular cutoff radii, respectively, of the red and green edge atoms.  Figures (g) and (h) show the energy change during existence transfer from the red edge atom to the green edge atom with all other atoms being constrained at existence 1. The curves are normalized with respect to the energy of the initial configuration shown in figures (a-f).  The figure shows that as the two atoms are separated from each other, the potential barrier is completely removed in the case of identical sites. When the sites are different, the energy decays monotonically towards the most stable site.}
    \label{fig:distances}
\end{figure*}

\begin{figure}[H]
    \centering
    \includegraphics[scale=1.20]{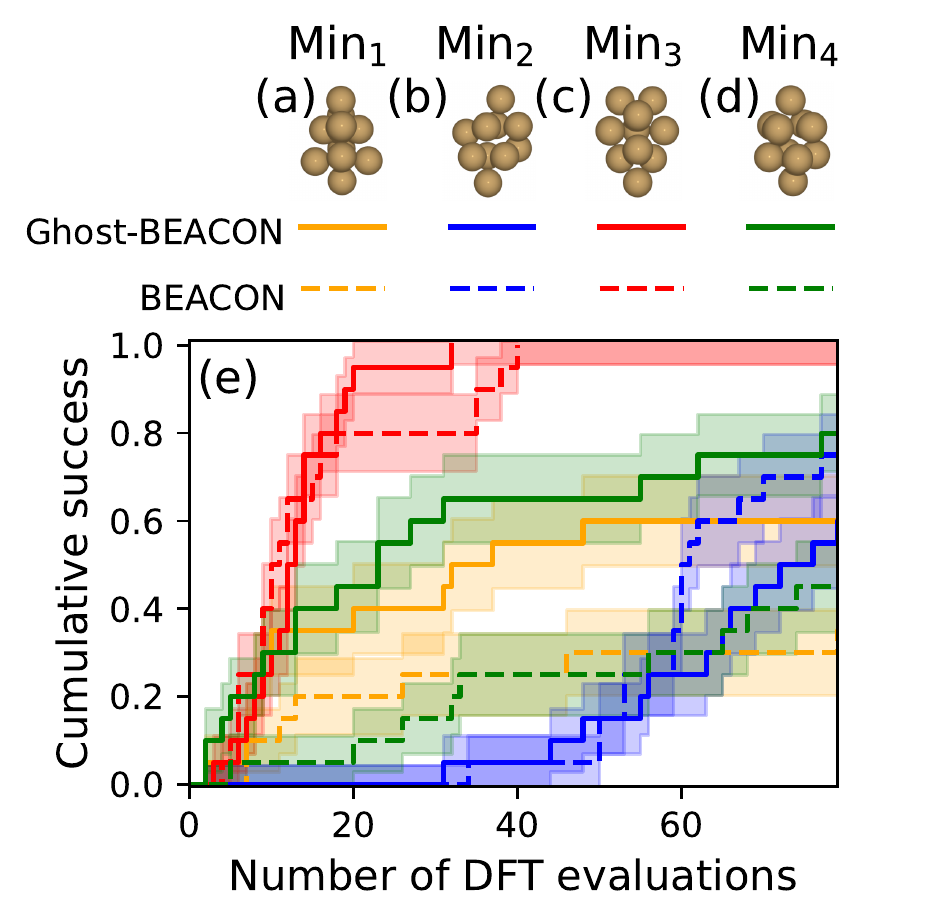}
    \caption{(a-d) The 4 lowest discovered minimum energy structures for Cu$_{10}$. (e): Success curves of 20 independent runs of each 80 DFT-calculations. Each iteration of the BEACON cycle is based on 40 surrogate relaxations. Each run has an initial training set of 2 random structures }
    \label{cluster_10}
\end{figure}

\begin{figure}[H]
    \centering
    \includegraphics[scale=0.90]{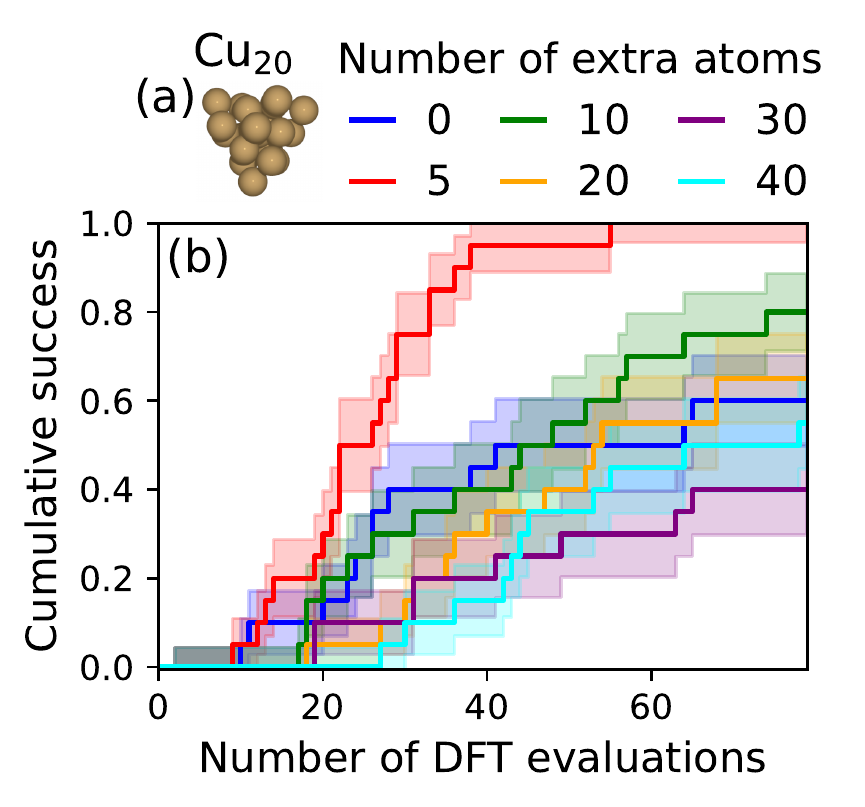}
    \includegraphics[scale=0.90]{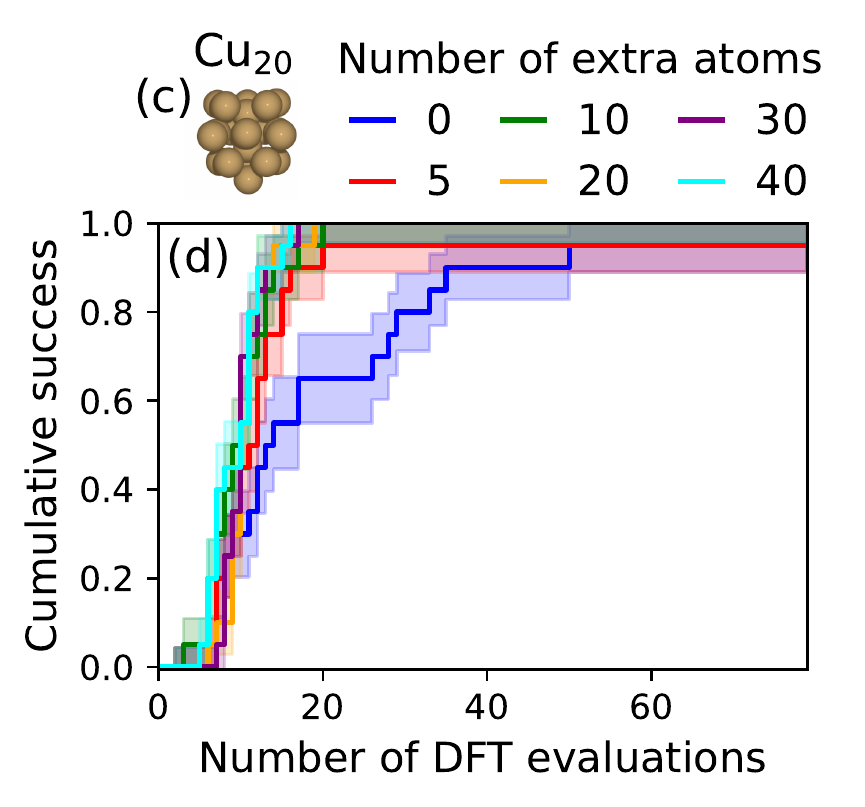}

    \caption{(a) Global minimum and (c) second lowest energy structure of Cu$_{20}$.  (b) and (d) Success-curves of 20 independent runs of each 80 DFT-calculations without ghost atoms (blue) and with 5 different numbers of ghost atoms for finding the structure shown in (a) and (c) respectively. Each iteration of the BEACON cycle is based on 40 surrogate relaxations. Each run has an initial training set of 2 random structures.} 
    \label{fig:cluster20}
\end{figure}

\begin{figure}[H]
    \centering
    \includegraphics[scale=0.8]{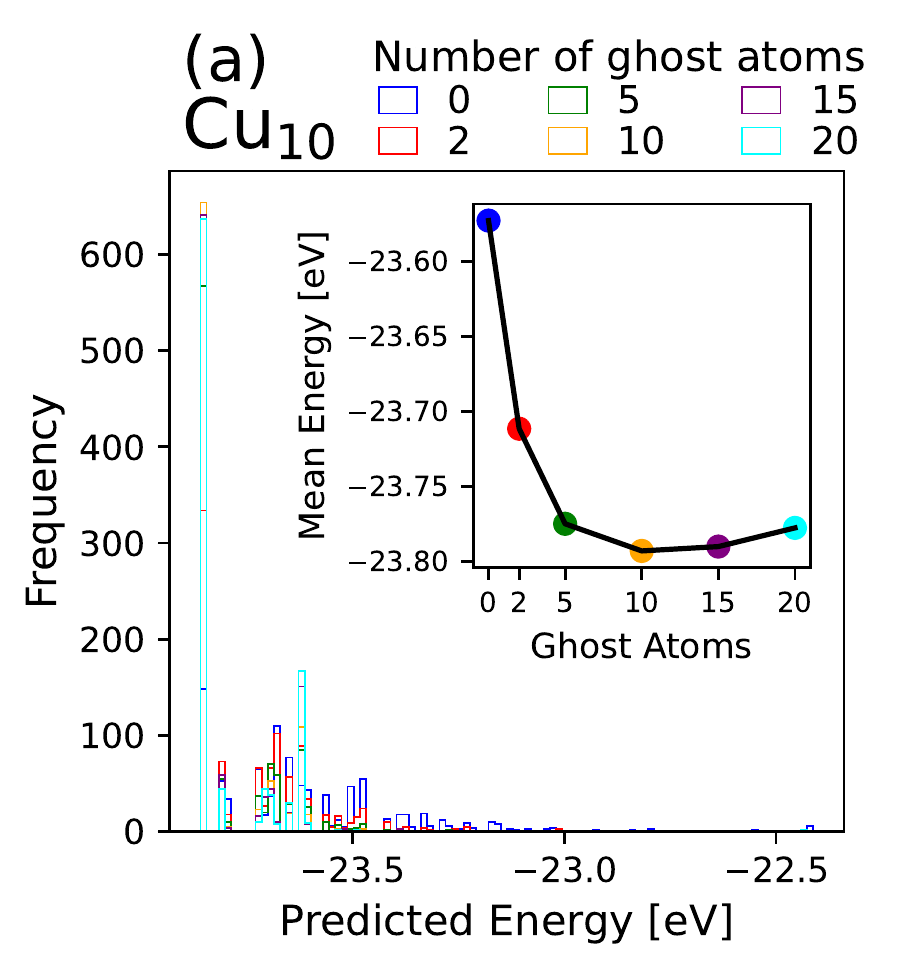}
    \includegraphics[scale=0.8]{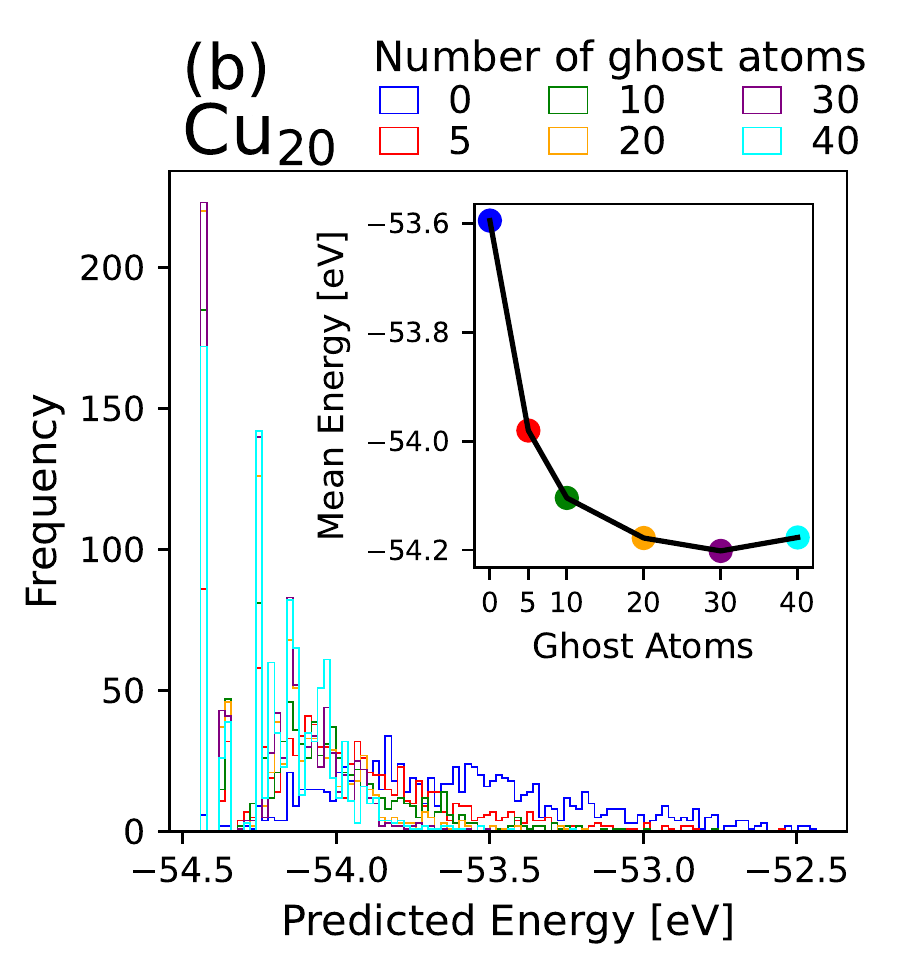}    
    \caption{Histograms of the predicted energies for 1000 surrogate relaxations of (a) Cu$_{10}$ and (b) Cu$_{20}$ for six different numbers of ghost atoms. The figures are similar to Fig.~5 in the main paper for Cu$_{30}$.}
    \label{fig:hists}
\end{figure}

\bibliography{supp_bib}